\documentclass[twocolumn]{aastex631}

\defcitealias{Davies21}{Paper~I}

\begin{document}

\title{\bf \large The Predicament of Absorption-dominated Reionization II: \\ Observational Estimate of the Clumping Factor at the End of Reionization}

\author[0000-0003-0821-3644]{Frederick B.~Davies}
\affiliation{Max-Planck-Institut f\"{u}r Astronomie, K\"{o}nigstuhl 17, D-69117 Heidelberg, Germany}

\author[0000-0001-8582-7012]{Sarah E.~I.~Bosman}
\affiliation{Institute for Theoretical Physics, Heidelberg University, Philosophenweg 12, D–69120, Heidelberg, Germany}
\affiliation{Max-Planck-Institut f\"{u}r Astronomie, K\"{o}nigstuhl 17, D-69117 Heidelberg, Germany}

\author[0000-0002-0658-1243]{Steven R.~Furlanetto}
\affiliation{Department of Physics \& Astronomy, University of California, Los Angeles, CA 90095, USA}

\begin{abstract}
    The history of reionization reflects the cumulative injection of ionizing photons by sources and the absorption of ionizing photons by sinks. The latter process is traditionally described in terms of a ``clumping factor'' which encodes the average quadratic increase in the recombination rate of dense gas within the cosmic web. The recent measurement of a short mean free path of ionizing photons from stacked quasar spectra at $z\simeq6$ has placed the importance of sinks under increased scrutiny, but its connection to the recombination rate is not immediately obvious. Here we present analytic arguments to connect the clumping factor to the mean free path by invoking ionization equilibrium within the ionized phase of the intergalactic medium at the end of (and after) reionization. We find that the latest mean free path and hydrogen photoionization rate measurements at $z=5$--$6$ imply a global clumping factor $C\approx12$, much higher than previous determinations from radiation-hydrodynamic simulations of the reionization process. Similar values of $C$ are also derived when applying the same procedure to observations at $2<z<5$. Compared to the traditional assumption of $C=3$, high-redshift galaxies must produce roughly twice as many ionizing photons ($\approx3$ photons per baryon) to reionize the universe by $z\sim6$. This additional requirement on the ionizing photon budget may help to reconcile the reionization history with JWST observations that suggest a far greater output of ionizing photons by the most distant galaxy populations.
\end{abstract}


\section{Introduction}

After the formation of baryons during the Big Bang, and their subsequent (re-)combination into atoms and the release of the cosmic microwave background (CMB), the hydrogen and helium in the Universe persisted in a predominantly neutral state. After the formation of the first stars and galaxies, the ionizing photons emitted by massive stars began to carve ionized bubbles into the surrounding intergalactic medium (IGM), beginning the epoch of reionization. The ionized bubbles from individual galaxies eventually merged \citep{Furlanetto04}, filling more and more of the cosmic volume until, by $z\sim5.3$ \citep{Bosman22}, the IGM was fully reionized.

The most straightforward quantitative description of reionization was put forward by \citet{Madau99}, whose ``one-zone'' model for the process provides valuable intuition:
\begin{equation}\label{eqn:m99}
    \frac{dQ}{dt} = \frac{\dot{n}_{\rm ion}}{\langle n_{\rm H} \rangle} - \frac{Q}{t_{\rm rec}},
\end{equation}
where $Q$ is the ionized fraction of the IGM, and the first and second terms on the right-hand side represent the source and sink terms, respectively. The sources are represented by $\dot{n}_{\rm ion}$, the emissivity of ionizing photons, while the sinks are represented by $t_{\rm rec}$, the recombination timescale of the ionized gas. In this work, as in \citet{Madau99}, we will assume that $Q$ represents the \emph{volume-averaged} ionized fraction. While the inside-out nature of reionization implies that a mass-averaged approach may be more appropriate, in which case an additional factor of $Q$ arises in the sink term (e.g.~\citealt{Chen20}), equation~(\ref{eqn:m99}) is approximately correct in a two-phase approximation (cf.~\citealt{Wu21b}) where the IGM is either fully ionized ($x_{\rm HII}=1$) or neutral ($x_{\rm HII}=0$). That is, $Q$ represents the volume of the IGM \emph{within the ionized phase} rather than a physical ionized fraction, and so it does not asymptote to a finite residual neutral fraction after reionization is complete (although see \citealt{Madau17} for a solution to this).

Considerable effort has been undertaken to determine $\dot{n}_{\rm ion}$ at early cosmic time. Such determinations typically involve a measurement of the UV luminosity function (LF) of galaxies at high redshifts ($z>6$), but the connection between the UV LF and the ionizing output of galaxies is still uncertain. This connection is usually parameterized as
\begin{equation}\label{eqn:nion}
    \dot{n}_{\rm ion} = \rho_{\rm UV} \xi_{\rm ion} f_{\rm esc},
\end{equation}
where $\rho_{\rm UV}$ is the integral over the UV LF down to some magnitude limit, $\xi_{\rm ion}$ is the ``ionizing efficiency'' which represents the average (intrinsic) spectral shape of the stellar populations between the ionizing and non-ionizing UV continuum, and $f_{\rm esc}$ is the escape fraction of ionizing photons from the galaxies into the IGM. While observations of galaxy nebular emission lines can constrain their $\xi_{\rm ion}$ (e.g.~\citealt{Ning23,Prieto-Lyon23,Simmonds23,Simmonds24,Atek24}), direct measurements of $f_{\rm esc}$ (i.e., direct detections of ionizing photons) are hindered by the high opacity of the Lyman-series forests, and thus are only possible at redshifts $z\lesssim4$ (e.g. \citealt{Izotov16,Vanzella18,Ji20,Pahl21}).

The role of sinks has been a subject of considerable debate over the years. As mentioned above, sinks of ionizing photons enter the \citet{Madau99} formalism via $t_{\rm rec}$, the average recombination time of a proton in the IGM, which can be written as
\begin{equation}\label{eqn:trec}
    t_{\rm rec} = \frac{1}{C \langle n_{\rm H} \rangle \alpha_{\rm HII}(T)},
\end{equation}
where $\alpha_{\rm HII}$ is the hydrogen recombination rate, and $\langle n_{\rm H} \rangle$ is the mean cosmic hydrogen density. Crucially, as recombination is a collisional process, the rate depends on the squared density of ionized gas. It is common to summarize this dependence with the so-called ``clumping factor,''
\begin{equation}\label{eqn:clump}
    C \equiv \langle n^2 \rangle / \langle n \rangle^2,
\end{equation}
such that $t_{\rm rec} = t_{\rm rec}^{\rm uniform} / C$. As there is no analytic shortcut to fully determine $C$ from first principles, the assumed value is typically derived from cosmological simulations.

Early cosmological hydrodynamical simulations suggested $C\sim30$ (e.g. \citealt{GO97}), implying a dominant role of recombinations in determining the reionization history. However, this high value considered all gas particles in the simulation volume. In practice, recombinations occurring \emph{inside} of galaxies are already accounted for by the $f_{\rm esc}$ parameter, so care must be taken to avoid double-counting. Later works by \citet{Iliev05,Iliev07} and \citet{Raicevic11} took this exclusion into account and found values closer to $C\sim10$, albeit using dark-matter-only N-body simulations. But the baryons are also subject to gas pressure, which smooths their distribution relative to the dark matter field. Following several works employing radiation hydrodynamic simulations of reionization \citep[][see also \citealt{McQuinn11}]{Pawlik09,Shull12,Finlator12}, a value of $C\sim2$--$3$ is now a typical assumption in analytic reionization models in the literature. That said, more recent simulations suggest a value of $C$ higher by a factor of $\sim2$ (e.g.~\citealt{Chen20,Kannan22}).

The impact of sinks is now being revisited after the measurement of a short mean free path of ionizing photons at $z=6$ by \citet{Becker21}, and further confirmed by \citet[][see also \citealt{Bosman21MFP,Satyavolu23}]{Zhu23}, which suggests the presence of substantially more small-scale structure in the IGM than present in traditional reionization simulations. A short mean free path can dramatically increase the number of ionizing photons required to ionize the IGM, as photons must travel long distances from ionizing sources to large-scale voids (\citealt{Davies21}, henceforth \citetalias{Davies21}, see also \citealt{Cain21}). However, the quantitative connection between the short mean free path and the clumping factor is not immediately obvious (e.g.~\citealt{Cain23}), which has limited the extent to which the short mean free path has been taken into account by the larger reionization community.

In this work, we aim to build a stronger connection between the mean free path and clumping factor at high redshift, in an attempt to unify the description of ionizing photon sinks during the reionization epoch. We first show that the clumping factor at a given redshift can be estimated from the ionizing background intensity and mean free path. We then apply this methodology to measurements of these quantities across cosmic time, finding a nearly constant value of $C$ which is several times higher than typically assumed.

We assume a \emph{Planck} $\Lambda$CDM cosmology \citep{Planck18} with $h=0.68$, $\Omega_m=0.31$, and $\Omega_b=0.049$.

\section{The clumping factor and the mean free path}\label{sec:clump}

In this section, we will investigate the connection between the clumping factor and the mean free path of ionizing photons. But first, we must define what we mean by ``clumping factor,'' as its exact definition varies considerably between different works. Here we define the clumping factor to be the relevant clumping factor for solving equation~(\ref{eqn:m99}) -- i.e.~the clumping factor that provides the correct globally averaged recombination rate, but where effects inside of galaxies that give rise to the escape fraction in the definition of $\dot{n}_{\rm ion}$ are ignored. Specifically, we assume that $C$ is the constant of proportionality between the true global (external) recombination rate $\dot{n}_{\rm rec} = \langle n_e n_{\rm HII} \alpha_{\rm HII} \rangle$ and the recombination rate at the cosmic mean density,
\begin{equation}\label{eqn:Crec}
    C \equiv \frac{\langle n_e n_{\rm HII} \alpha_{\rm HII} \rangle }{\langle n_e \rangle \langle n_{\rm HII} \rangle \langle \alpha_{\rm HII} \rangle}
\end{equation}
where we assume that the fiducial recombination coefficient in the denominator is equal to the Case B recombination rate for $10,000$\,K gas\footnote{We note that $\alpha_{\rm HII}^{B}(T=10,000\,{\rm K})\simeq\alpha_{\rm HII}^{A}(T=20,000\,{\rm K})$, another common assumption in previous works.}, $\langle \alpha_{\rm HII} \rangle=\alpha_{\rm HII}^{B}(T=10,000\,{\rm K})$.

The most straightforward way to connect the clumping factor to the mean free path starts with the assumption of photoionization equilibrium, which should generally hold in the ionized IGM,
\begin{equation}
    n_{\rm HI} \Gamma_{\rm HI} = \langle n_e n_{\rm HII} \alpha_{\rm HII} \rangle = C \chi_e  (1-x_{\rm HI})^2 n_{\rm H}^2 \alpha_{\rm HII},
\end{equation}
where $\Gamma_{\rm HI}$ is the photoionization rate of hydrogen and $\chi_e\approx1.08$ is the enhancement in the number of free electrons due to ionized helium\footnote{We assume that helium is singly-ionized at the same time as hydrogen, and that the (second) reionization of helium has not yet begun. While this assumption will become incorrect at $z\lesssim4$ (e.g. \citealt{Worseck19}), the additional 8\% boost to the electron density from the second ionization of helium is small compared to the uncertainties in the observed quantities we employ in \S~\ref{sec:measure}.}. Solving for $C$, we have
\begin{equation}\label{eqn:7}
    C = \frac{x_{\rm HI} n_{\rm H} \Gamma_{\rm HI}}{\chi_e (1-x_{\rm HI})^2 n_{\rm H}^2 \alpha_{\rm HII}}.
\end{equation}
The two unknowns in this expression are $\Gamma_{\rm HI}$ and $x_{\rm HI}$. While the former can be derived from observations of the Ly$\alpha$ forest, and is most sensitive to low-density gas which is unambiguously resolved in simulations, the residual neutral fraction is not so simple to derive, as it is sensitive to self-shielding and geometrical effects in dense gas (e.g.~\citealt{McQuinn11,Erkal15}).

To proceed, one can make the simplifying assumption that the neutral fraction is connected to the mean free path of ionizing photons via $\lambda_{\rm mfp} = (n_{\rm HI} \sigma_{\rm HI})^{-1}$. In this case, similar to the ``effective'' clumping factor in \citet{Cain23},  we have:
\begin{equation}\label{eqn:Cfac1}
    C = \frac{\sigma_{\rm HI}\Gamma_{\rm HI}}{\lambda_{\rm mfp}\chi_e (1-x_{\rm HI})^2 n_{\rm H}^2 \alpha_{\rm HII}},
\end{equation}
where we note that both the mean free path and photoionization cross section terms implicitly represent frequency-averaged values. This expression for the mean free path, however, makes the crucial assumption that neutral hydrogen is uniformly distributed in space. In reality, the dense self-shielded gas that gives rise to optically-thick absorption should be inhomogeneous, distributed in clumps and/or in the filaments of the cosmic web (e.g.~\citealt{McQuinn11}).

Another way to associate the clumping factor with the mean free path was suggested by \citet{Emberson13}, who connected the attenuation of ionizing photon flux to the recombination rate. In this model, one considers the attenuation of ionizing photon flux $dF$ inside a slab of material with area $dA$ and proper width $ds$ compared to the recombinations inside said slab,
\begin{equation}
    -dF\,dA = C n_e n_{\rm HII} \alpha_{\rm HII}\,dA\,ds,
\end{equation}
where the flux inside the slab is attenuated following
\begin{equation}
    \frac{dF}{ds} = -F(1+z)/\lambda_{\rm mfp},
\end{equation}
where the $(1+z)$ term converts $\lambda_{\rm mfp}$ from comoving to proper units.
The following expression can then be derived after solving for $C$,
\begin{equation}\label{eqn:Cfac2}
    C = \frac{F(1+z)}{\lambda_{\rm mfp} \chi_e (1-x_{\rm HI})^2 n_{\rm H}^2 \alpha_{\rm HII}},
\end{equation}
where we note again that the $F$ and $\lambda_{\rm mfp}$ terms represent values integrated over the spectrum of the ionizing background. While this expression improves upon the previous one by removing the explicit connection between the mean free path and the neutral fraction, the geometrical assumption in its derivation (i.e.~the slab) introduces additional ambiguity.

We suggest a third way to conceptualize (and quantify) the connection between the mean free path and the clumping factor. A typical ionizing photon passing through the IGM will travel one mean free path before being absorbed, i.e.~before ionizing a hydrogen atom. Thus a ``photon photo-ionization rate,'' the rate at which a given ionizing photon will ionize a hydrogen atom, can be written as $\Gamma_\gamma = c/\lambda_{\rm mfp}$. The space density of photoionizations is then $n_\gamma \Gamma_\gamma = n_\gamma\times c/\lambda_{\rm mfp}$, where $n_\gamma$ is the number density of ionizing photons. In ionization equilibrium, this rate will balance the recombination rate, i.e.
\begin{equation}
    \frac{n_\gamma c}{\lambda_{\rm mfp}} = C \alpha_{\rm HII} \chi_e (1-x_{\rm HI})^2 n_H^2,
\end{equation}
Solving for $C$ as above, we find
\begin{equation}\label{eqn:Cfac3}
    C = \frac{n_\gamma c}{\lambda_{\rm mfp} \alpha_{\rm HII} \chi_e (1-x_{\rm HI})^2 n_{\rm H}^2},
\end{equation}
where again the $n_\gamma$ and $\lambda_{\rm mfp}$ terms represent frequency-averaged quantities.

In all three cases, the inferred clumping factor is proportional to the strength of the ionizing background (in various forms) divided by the mean free path of ionizing photons, e.g.~$C\propto \Gamma_{\rm HI}/\lambda_{\rm mfp}$. All three methods result in quantitatively similar values for $C$; in this work, we adopt the third method to compute $C$ (i.e.~equation~\ref{eqn:Cfac3}), as it appears to have the fewest explicit assumptions on the nature of the distribution of neutral gas.

We have so far ignored the dependence on photon frequency of various quantities in the expressions for $C$ above for the sake of clarity, but due to the steep frequency dependence of the photoionization cross-section \citep{Verner96}, such terms could matter at the level of a factor of a few. We write the specific number density of ionizing photons $n_\nu$ as
\begin{equation}
    n_\nu = \frac{u_\nu}{h\nu} = \frac{4\pi}{c} \frac{J_\nu}{h\nu},
\end{equation}
where $u_\nu$ is the specific energy density and $J_\nu$ is the specific angle-averaged mean intensity of the ionizing background. We then proceed to estimate $C$ using the following expression:
\begin{equation}\label{eqn:Cfid}
    C = \left[4\pi\int_{\nu_{\rm HI}}^{4\,\nu_{\rm HI}} \frac{J_\nu}{h\nu\lambda_\nu} {\rm d}\nu \right]\times\frac{1}{\alpha_{\rm HII} \chi_e (1-x_{\rm HI})^2 n_{\rm H}^2},
\end{equation}
where $\nu_{\rm HI}$ is the frequency of the hydrogen ionizing edge. In the following, we make the assumption that the neutral fraction is small enough that the $(1-x_{\rm HI})$ term can be approximated as unity -- that is, we compute the clumping factor relative to a fully ionized IGM. We further approximate the frequency dependencies of the mean free path and ionizing background intensity as power laws with $\lambda_\nu\propto\nu^{\alpha_\lambda}$ and $J_\nu\propto\nu^{-\alpha_{\rm b}}$, leading to an analytic simplification to equation~(\ref{eqn:Cfid}) as long as $\alpha_{\rm b}+\alpha_\lambda\neq0$,
\begin{equation}\label{eqn:Cfid2}
    C = \frac{4\pi J_{\rm HI}}{h\nu_{\rm HI}\lambda_{\rm mfp}}\left[\frac{1-4^{-(\alpha_{\rm b}+\alpha_\lambda)}}{\alpha_{\rm b}+\alpha_\lambda}\right] \times \frac{1}{\alpha_{\rm HII}\chi_e n_{\rm H}^2},
\end{equation}
where $J_{\rm HI}$ and $\lambda_{\rm mfp}$ are $J_\nu(\nu_{\rm HI})$ and $\lambda_\nu(\nu_{\rm HI})$, respectively.

The frequency dependence assumptions are largely encapsulated by the term in brackets, which is a function of $\alpha_b+\alpha_\lambda$, equal to $2$ with our assumed power-law indices\footnote{We note that at $z\gtrsim5$, where the mean free path is short relative to the Hubble distance (the ``absorption-limited'' regime), we can write $J_\nu\propto\epsilon_\nu\lambda_\nu$ where $\epsilon_\nu$ is the average ionizing emissivity \citep{MW03}. The implied scaling of the emissivity is $\epsilon_\nu\propto\nu^{-(\alpha_b+\alpha_\lambda)}$, suggesting that our choice of $\alpha_b+\alpha_\lambda=2$ is reasonable (e.g.~\citealt{BB13}). Even harder ionizing spectra are also plausible for young, metal-poor stellar populations (e.g.~\citealt{D'Aloisio19}) which would increase our $C$ estimates.}. In practice, as described below in \S~\ref{sec:measure}, we will convert observational constraints on $\Gamma_{\rm HI}$ to $J_{\rm HI}$, which introduces an additional factor dependent on $\alpha_b$ alone. Lower (higher) values of $C$ would be derived if the ionizing background is softer (harder), or if the mean free path is a stronger (weaker) function of frequency. We discuss the behavior of the frequency dependence term further in Appendix~\ref{sec:freq}, and note that even in the most extreme case (corresponding to $\alpha_\lambda\approx3$) our assumptions could only overestimate $C$ by a factor of two.

As measurements of the ionizing background require non-zero transmission through the highly sensitive Ly$\alpha$ forest \citep{GP65}, application of this method will only become possible at the very end of the reionization process. Thus the clumping factor during the majority of reionization cannot be constrained directly. At these earlier times, the inside-out nature of reionization should lead to higher densities in the ionized regions, and thus a higher recombination rate relative to the mean IGM (e.g.~\citealt{Chen20}, see also \hbox{\citealt{So14})}. We ignore this effect for simplicity in our analytic calculations, but note that this would likely increase the effective clumping factor significantly at low ionized fractions as in \citet{Chen20}.

\section{Estimating the clumping factor from IGM observations}

In this section, we use observational properties of the IGM to constrain the clumping factor as defined in the previous section. Here we stress that our clumping factor is an effective quantity corresponding to the entire volume of the IGM, and not a ``local'' clumping factor that can be applied to the density field in simulations a posteriori (see, e.g., \citealt{Raicevic11,KG15}). In particular, our definition of the clumping factor is specifically designed for use in analytic reionization calculations like equation~(\ref{eqn:m99}) that consider the IGM as a whole \citep{Madau99}. 

\begin{figure}[t]
\begin{center}
\resizebox{8.5cm}{!}{\includegraphics[trim={1em 1em 1em 1em},clip]{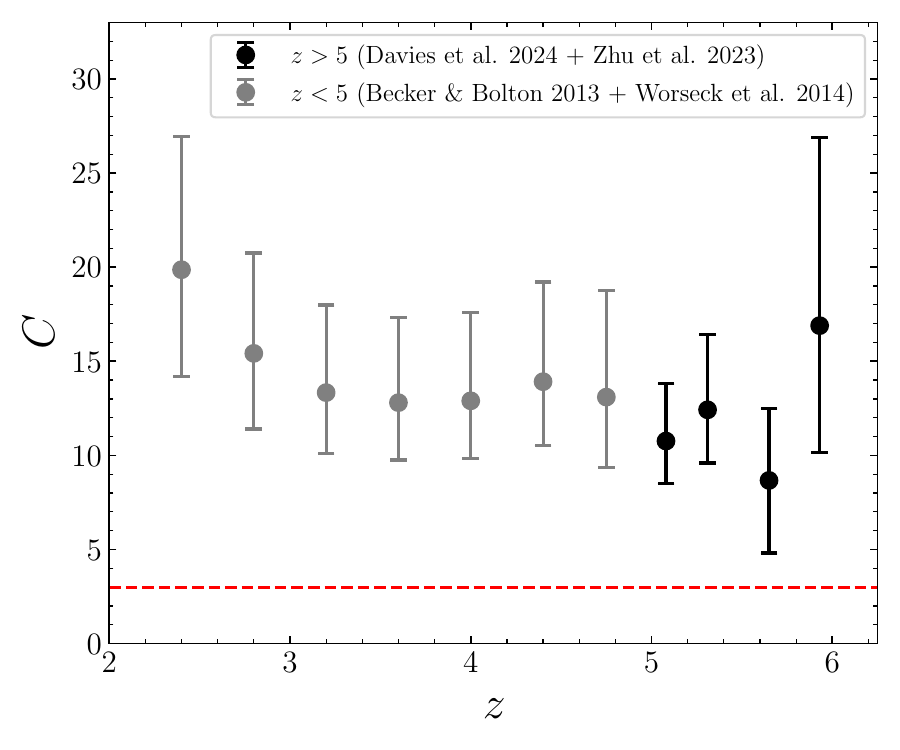}}\\
\end{center}
\caption{Clumping factor estimates applying equation~(\ref{eqn:Cfid2}) to constraints on the hydrogen photoionization rate $\Gamma_{\rm HI}$ (via equation~\ref{eqn:GJ}) and ionizing photon mean free path $\lambda_{\rm mfp}$. The grey points at $z<5$ use $\Gamma_{\rm HI}$ from \citet{BB13} and $\lambda_{\rm mfp}$ from \citet{Worseck14}, while the black points at $z>5$ use the corresponding constraints from \citet{Davies24} and \citet{Zhu23}, respectively.}
\label{fig:measure}
\end{figure}

\begin{figure*}[ht]
\begin{center}
\resizebox{16cm}{!}{\includegraphics[trim={1em 1em 1em 1em},clip]{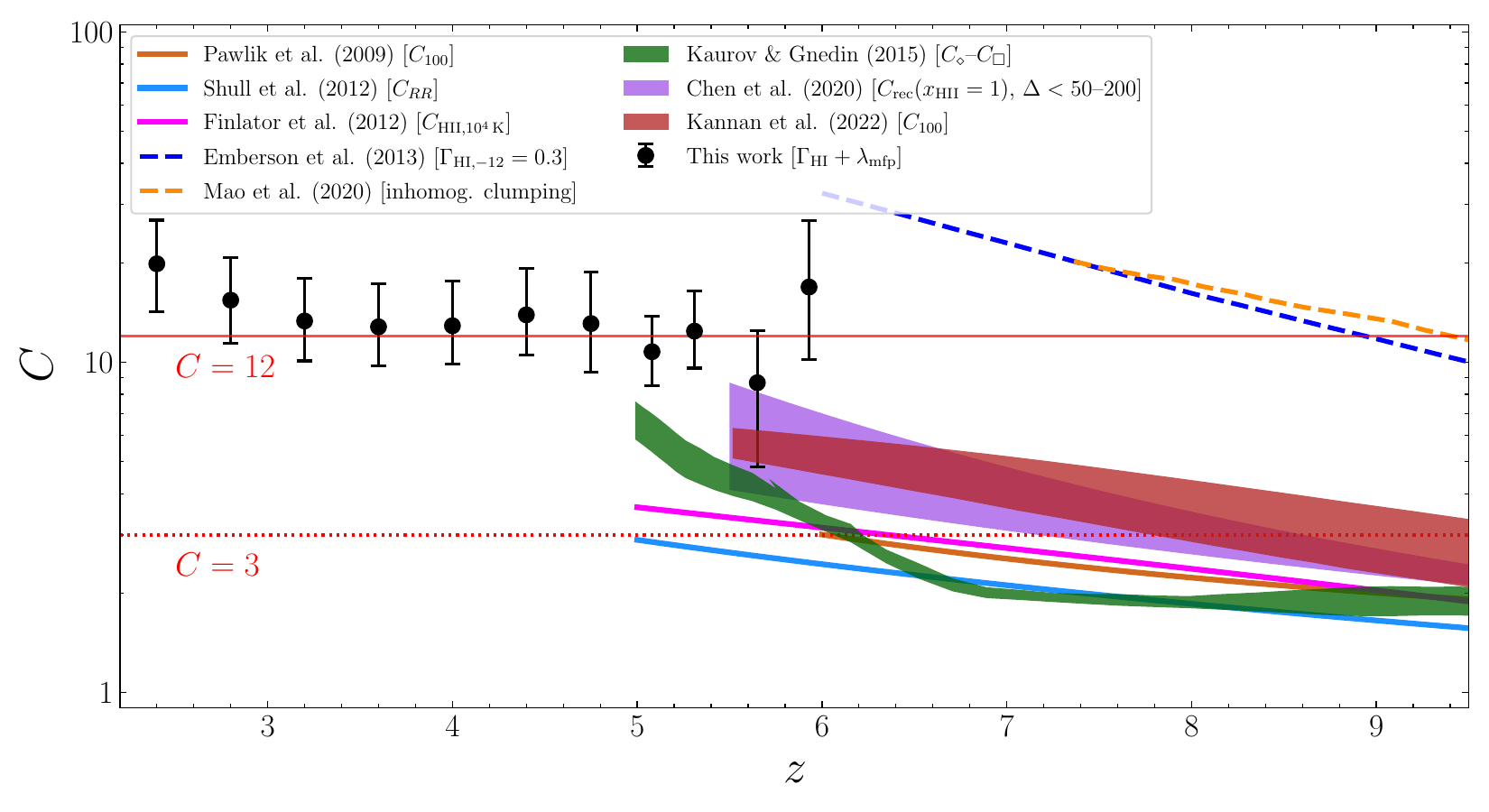}}\\
\end{center}
\caption{Comparison between the $C$ estimates from this work (black points) and from various cosmological simulations. The solid curves show fits to the simulations from \citet{Pawlik09} in brown, \citet{Shull12} in light blue, and \citet{Finlator12} in pink. The shaded regions show ranges in $C$ estimates from \citet{KG15} in green and \citet{Chen20} in purple, where for the latter we have set the ionized fraction to unity to mimic our assumptions. The dashed curves show simulations without photoionization heating, with the small-box adiabatic hydrodynamical simulation from \citet{Emberson13} in blue and the N-body simulations from \citet{Mao20} in orange.}
\label{fig:compare}
\end{figure*}

\subsection{Observations of $\Gamma_{\rm HI}$ and $\lambda_{\rm mfp}$}\label{sec:obs}

Estimating the clumping factor using the method above requires an estimate of the ionizing background intensity as well as the mean free path of ionizing photons. While most studies of the clumping factor have focused on its behavior during the reionization epoch, our formalism applies to any cosmic time where both of these quantities have been measured. 

For the mean free path, at $z<5$ we use the power-law fit to the direct measurements of quasar spectra stacked beyond the Lyman limit from \citet{Worseck14} and references therein. At $z>5$, we use the measurements from \citet{Zhu23}, who (following \citealt{Becker21}) use a similar stacking method to \citet{Worseck14} but additionally account for the bias due to the intense local ionizing flux from the background quasars. At all redshifts we assume a power-law frequency dependence of the mean free path of $\lambda_\nu \propto \nu^\alpha_\lambda$, with $\alpha_\lambda=1$. This power-law dependence approximately corresponds to a \ion{H}{1} column density distribution function proportional to $N^{-4/3}$. This distribution shape is somewhat flatter than the distribution of lower density Ly$\alpha$ forest absorbers, but consistent with some measurements and models at $z=2$--$6$ \citep{FG09,SC10}. We note that this connection to the column density distribution is only an approximation, as the shape is likely much more complicated in the relevant range of \ion{H}{1} column densities (e.g.~\citealt{HM12,O'Meara13,Prochaska14}).

For the ionizing background, at $z<5$ we adopt the measurements of $\Gamma_{\rm HI}$ from \citet{BB13}, who calibrated a suite of hydrodynamical simulations to the mean transmitted flux of Ly$\alpha$ derived from a stacking analysis of SDSS quasars \citep{Becker13}, and comprehensively accounted for various sources of systematic uncertainty. At $z>5$, we use the constraints on $\Gamma_{\rm HI}$ from \citet{Davies24}, who fit a fluctuating ionizing background model \citep{DF16} to the Ly$\alpha$ forest opacity distributions from \citet{Bosman22}. To determine the specific intensity at the hydrogen-ionizing edge $J_{\rm HI}$ required by equation~(\ref{eqn:Cfid2}), we assume that the spectrum of the hydrogen-ionizing background (i.e.~$\nu_{\rm HI} < \nu < 4\,\nu_{\rm HI}$) is described by a power-law shape $J_\nu\propto\nu^{-\alpha_b}$ with $\alpha_b=1.0$. This spectral shape is consistent with an intrinsic ionizing emissivity proportional to $\nu^{-2}$ (cf.~\citealt{BB13,D'Aloisio19}) filtered through the absorber distribution giving rise to $\lambda_\nu\propto\nu^1$ as assumed above. We then determine the corresponding $J_{\rm HI}$ by requiring that the observed $\Gamma_{\rm HI}$ is reproduced by
\begin{equation}\label{eqn:GJ}
    \Gamma_{\rm HI} = 4\pi\int_{\nu_{\rm HI}}^{4\,\nu_{\rm HI}} \frac{J_{\rm HI} (\nu/\nu_{\rm HI})^{-\alpha_b}}{h\nu}\sigma_{\rm HI}(\nu) {\rm d}\nu,
\end{equation}
where $\sigma_{\rm HI}(\nu)$ is the hydrogen photoionization cross-section from \citet{Verner96}.

We note that the mean free path measurements from \citet{Zhu23} are derived assuming specific values of the hydrogen photoionization rate (and its uncertainty) from \citet{Gaikwad23}. To ensure self-consistency, we recompute the mean free path and corresponding uncertainties using the \citet{Davies24} constraints on $\Gamma_{\rm HI}(z)$, but note that this does not make a substantial difference to our results. 

\subsection{Estimates of the effective clumping factor}\label{sec:measure}

With the ionizing background strength and mean free path in hand, we can now proceed to compute the clumping factor following Section~\ref{sec:clump}. Specifically, we evaluate equation~(\ref{eqn:Cfid2}) using the mean free path measured by \citet{Worseck14} (i.e. the power-law fit from $z=2$--$5$) and \citet{Zhu23}, and the ionizing background intensity implied by the photoionization rate measurements of \citet{BB13} and \citet{Davies24}, with assumed frequency dependencies $\lambda_\nu \propto \nu$ and $J_\nu\propto\nu^{-1}$.

We show the resulting estimates of $C$ from $z=2$--$6$ in Figure~\ref{fig:measure}. The error bars at $z<5$ include only the uncertainty in $\Gamma_{\rm HI}$ from \citet{BB13}, while at $z>5$ they include both the uncertainty in $\Gamma_{\rm HI}$ from \citet{Davies24} and in $\lambda_{\rm mfp}$ from \citet{Zhu23}. We find a remarkably constant value of $C\sim10$--$15$ across the entire redshift range, with an average value of $C\approx12$ at $z=5$--$6$, well above the simulation-calibrated prescriptions often used in the literature ($C\sim3$). While the short mean free path at $z=6$ suggests a rather high value $C\sim17$, its corresponding uncertainty is large enough to be consistent with all lower redshifts.

We note that at $z\lesssim4$ we expect that our assumption of local photoionization equilibrium becomes increasingly incorrect. The mean free path at this time is long enough that the local source approximation is no longer valid (see, e.g., the discussion in \citealt{BB13}), so the photons being absorbed at a given epoch were emitted at a substantially earlier time. In addition, as the mean free path increases an increasing fraction of ionizing photons will redshift below $\nu_{\rm HI}$ before encountering a hydrogen atom. Neglecting these effects is likely the cause for the upturn in our $C$ estimates at $z<3$ visible in Figure~\ref{fig:measure}.

\subsection{Comparison to simulations}

In Figure~\ref{fig:compare}, we compare our estimates of $C$ to various determinations of $C$-like quantities in the literature. The solid curves from \citet{Pawlik09}, \citet{Shull12}, and \citet{Finlator12} represent the basis behind the commonly-assumed values of $C=2$--$3$, with the ranges of estimates from more recent simulations by \citet{KG15}, \citet{Chen20}, and \citet{Kannan22} shown as shaded regions with moderately higher values up to $C\sim5$ at $z=5$--$6$. All of these works compute the clumping factor in different ways, but in principle they are all designed to fulfill the same role: to quantify the effect of the sink term on the progression of reionization in equation~(\ref{eqn:m99}).

The dashed curves in Figure~\ref{fig:compare} show clumping factors measured from simulations without any contribution from photoionization heating, with \citet{Emberson13} employing small-volume adiabatic hydrodynamical simulations and \citet[][see also \citealt{Bianco21}]{Mao20} using a combination of small and large N-body simulations. These curves can be considered to be theoretical ``maximum'' values for $C$, and our estimates lie comfortably below them.

Why, then, do we recover such a large value for $C$ compared to the commonly-accepted value from simulations? In the absence of an unforeseen source of bias in our approach, it is possible that the $C$ measured in simulations is not directly comparable to our value of $C$ due to a difference in definition. Simulations are careful to compute $C$ without including dense gas inside of halos, to avoid double-counting this gas which could be responsible for the galactic escape fraction. This is typically implemented as a density threshold, e.g. $C_{100}$ from \citet{Pawlik09} is measured from gas with overdensity $\Delta < 100$. Other works include cuts on the temperature and ionization state of the gas (e.g. \citealt{Finlator12,KG15}).

\begin{figure}[t]
\begin{center}
\resizebox{8.5cm}{!}{\includegraphics[trim={1em 1em 1em 1em},clip]{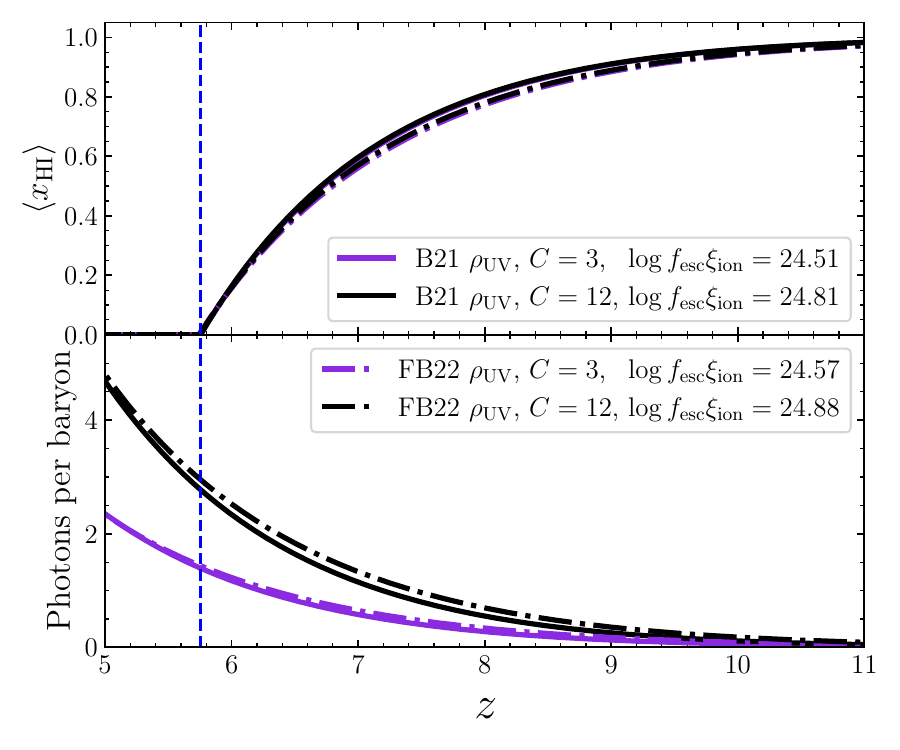}}\\
\end{center}
\caption{Reionization histories (upper panel) computed using equation~(\ref{eqn:m99}) with $C=3$ (purple) and $C=12$ (black) and the cumulative number of ionizing photons per baryon (lower panel). We integrate the UV LFs from \citet[][solid]{Bouwens21} and \citet[][dot-dashed]{FB22} down to $M_{\rm UV}=-13$ to compute the UV luminosity density, and multiply by fixed values of $f_{\rm esc}\xi_{\rm ion}$ given in the legend to calibrate $\dot{n}_{\rm ion}$ to reach $Q=0.9$ at $z=5.9$, which then finishes reionization at $z\approx5.7$ (vertical dashed line). While our estimate of $C=12$ does not affect the reionization history compared to the classical assumption of $C=3$ in this scenario, it requires a factor $\sim2$ higher total photon output from galaxies to finish reionization.}
\label{fig:history}
\end{figure}

But this dense gas masking ignores the crucial possibility that dense halo gas can also be illuminated from the \emph{outside} by the UV background, and potentially make up a substantial fraction of the opacity to ionizing photons streaming through the IGM. The mean free path measured in stacked quasar spectra (e.g.~\citealt{Prochaska09,Worseck14,Becker21,Zhu23}) or from the \ion{H}{1} column density distribution \citep{Rudie13,Prochaska14}, includes encounters of ionizing photons with all gas without any regard for whether it is associated with a galaxy.

Recent simulations by \citet{Cain23} which take into account the effect of IGM small-scale ($\sim$\,kpc) structure and its relaxation dynamics after reionization heating \citep{Park16,D'Aloisio20,Chan24}, have found that applying a clumping factor $C=5$ to their coarse 1\,Mpc-resolution simulation provides a decent match to their more sophisticated sink modeling. On the surface, this value is substantially lower than our estimates, but recall that our $C$ is defined \emph{globally} -- this distinction is important, because the locally-defined $C$ can be much smaller than the global one \citep{Raicevic11}. In fact, the scale-dependence of the clumping factor found by \citet{KG15} suggests that the global $C$ is $\sim2$ times larger than the local $C$ on $\sim1$\,Mpc scales, implying that our estimate for the (global) $C$ is reasonably consistent with the model from \citet{Cain23}.

\section{Implications for reionization}\label{sec:discuss}

Fundamentally, the purpose of estimating this particular definition of the clumping factor is to explore what implications the short mean free path of ionizing photons at $z=6$ has for the reionization history, and particularly, for the requirements on the number of ionizing photons that must have been emitted to complete the process. The semi-numerical simulations in \citetalias{Davies21} examined this in the context of the way a short mean free path inhibits the ionization of the last remaining voids; here, instead, we consider solely the effect of the additional recombinations inside of ionized gas from the large clumping factor implied by IGM observations at $z=5$--$6$ as shown above. While analytically convenient, this choice comes at the expense of neglecting the effect of the spatial offset between ionizing sources and the last patches of neutral gas at the end of reionization (\citetalias{Davies21}; \citealt{DF22}); we leave a closer look at that effect to future work. In this section we will consider the impact of our high value of $C=12$ on reionization calculations involving equation~(\ref{eqn:m99}).

\begin{figure}[t]
\begin{center}
\resizebox{8.5cm}{!}{\includegraphics[trim={1em 1em 1em 1em},clip]{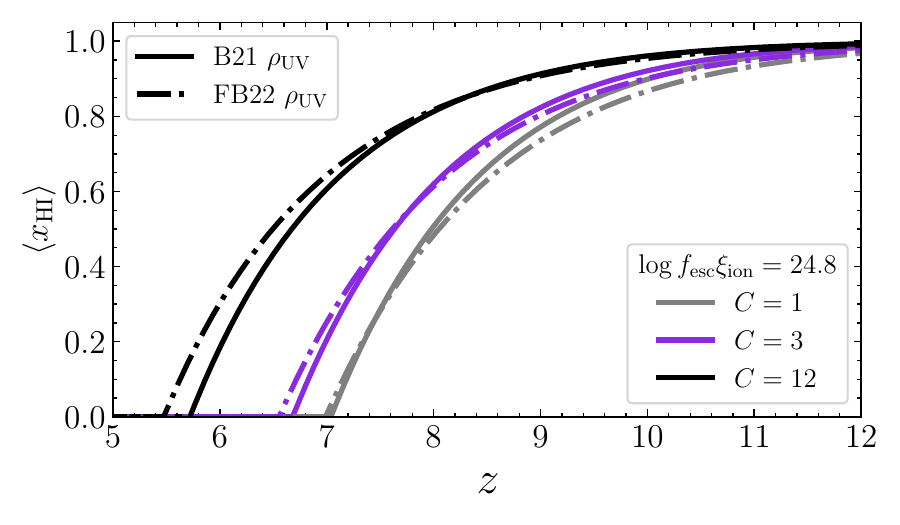}}\\
\end{center}
\caption{Similar to the upper panel of Figure~\ref{fig:history} but keeping $f_{\rm esc}\xi_{\rm ion}$ fixed to $10^{24.8}$\,erg/Hz, and including the case of a uniform IGM ($C=1$, grey).}
\label{fig:history1}
\end{figure}

\begin{figure*}[t]
\begin{center}
\resizebox{18cm}{!}{\includegraphics[trim={1em 1em 1em 1em},clip]{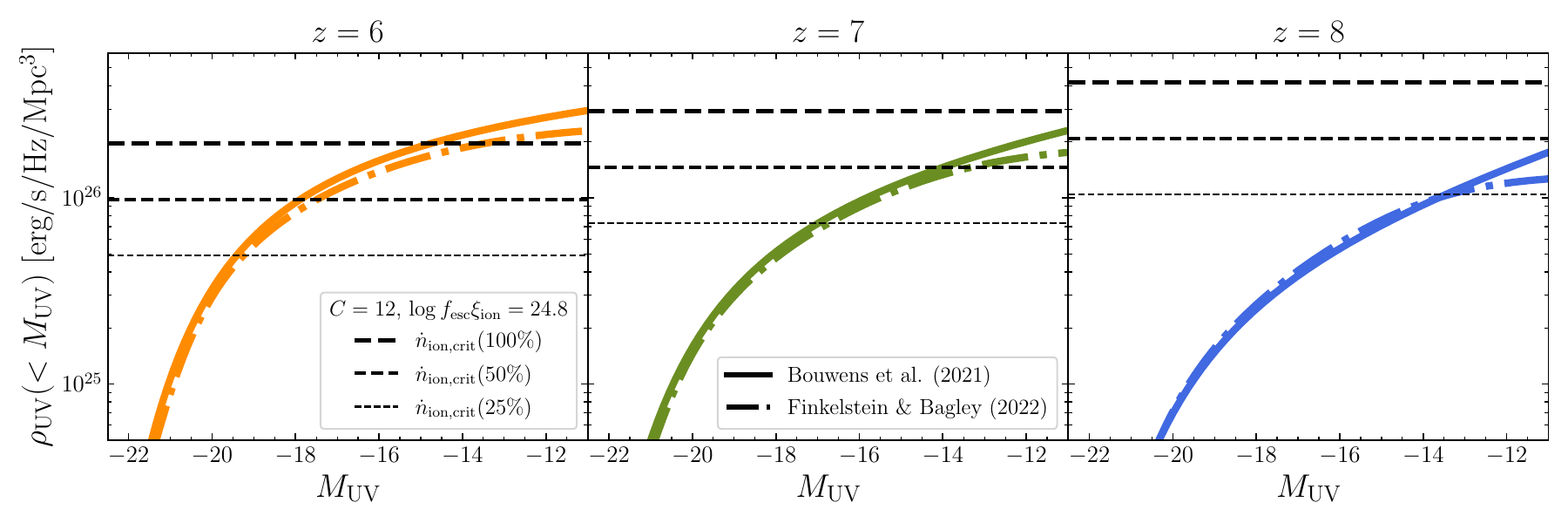}}\\
\end{center}
\caption{Required UV luminosity density from galaxies in order to maintain the Universe fully, $50\%$, and $25\%$ ionized (thick to thin dashed lines). Assuming an effective ionization production $f_{\text{esc}}\xi_{\text{ion}}=10^{24.8}$\,erg/Hz and a clumping factor $C=12$, the integrated light from galaxies down to $M_{\text{UV}}=-13$ is capable of maintaining a fully ionized the Universe at $z=6$, but only $<50\%$ at $z=8$. The results depend only weakly on the choice of UV LF, \citet{Bouwens21} (solid) or \citet{FB22} (long-dashed).}
\label{fig:nioncrit}
\end{figure*}

We must first consider our model for the sources, i.e.~$\dot{n}_{\rm ion}$ (equation~\ref{eqn:nion}). We compute $\rho_{\rm UV}(z)$ from the evolving UV LF parameterizations by \citet{Bouwens21} and \citet{FB22}, extrapolating up to $z=15$, and integrating down to a fiducial limiting UV magnitude of $M_{\rm UV}=-13$. The redshift evolution of the \citet{FB22} LF includes a strongly evolving suppression at the faint end leading to a rapid decline in the UV luminosity density at the upper end of (and beyond) their fitting range at $z\gtrsim9$. We thus extrapolate to higher redshift with a double-power-law fit to the evolution at $5<z<8.5$, although we note that this makes little difference to our main results.


We can now explore the consequences for the reionization history, integrating equation~(\ref{eqn:m99}) across cosmic time. We adopt clumping factors of $C=3$, representing the traditional approach, and $C=12$, as determined in this work. We then tune the product of the ionizing escape fraction and ionizing efficiency $f_{\rm esc}\xi_{\rm ion}$ in each case to reach a neutral fraction of $10\%$ at $z=5.9$, consistent with the Ly$\alpha$ forest dark pixel constraint from \citet{McGreer15} and with the model from \citetalias{Davies21}, leading to a late end to reionization consistent with the most recent constraints from the Ly$\alpha$ forest \citep{Zhu21,Zhu22,Zhu24,Bosman22,Spina24}. 

The resulting reionization histories are shown in the top panel of Figure~\ref{fig:history}. At this fixed endpoint of reionization, and with our particular models for $\dot{n}_{\rm ion}(z)$, increasing the clumping factor from $C=3$ to $C=12$ has a negligible effect on the reionization history at earlier times. In the lower panel of Figure~\ref{fig:history}, we show the corresponding integrated number of ionizing photons per baryon. Assuming $C=3$ requires $1.3$--$1.5$ photons per baryon to complete reionization, while with $C=12$ the number doubles to $2.5$--$3.0$ photons per baryon. This elevated photon budget is nevertheless still slightly below the nominal range from \citetalias{Davies21}, but it is very similar to the dynamic sink radiative transfer models of \citet{Cain21}. This number is also consistent with the total number of recombinations at the end of reionization (i.e. the number of emitted photons per baryon minus one) in the CROC radiation hydrodynamical simulations \citep{Gnedin22,Gnedin24}.

More generally, if we remove the restriction on the endpoint of reionization, we can explore how different values of $C$ change the reionization history at fixed $\dot{n}_{\rm ion}(z)$. In Figure~\ref{fig:history1}, we show reionization histories similar to Figure~\ref{fig:history} but with a fixed $f_{\rm esc}\xi_{\rm ion}=10^{24.8}$\,erg/Hz, corresponding to e.g.~a model with $\xi_{\rm ion}=10^{25.8}$\,erg/Hz, consistent with a recent determination for UV-faint galaxies with JWST \citep{Atek24}, and $f_{\rm esc}=0.1$, consistent with direct measurements of Lyman continuum photons from Lyman-break galaxies at $z\sim3$ \citep{Pahl21}. While the conventional assumption of $C=3$ only modestly postpones the end of reionization by $\Delta z\sim0.3$ relative to a uniform IGM ($C=1$), our fiducial $C=12$ delays its completion by $\Delta z \gtrsim 1$.

Next, we examine the commonly-used criterion for reionization to remain complete, defined by setting $Q=1$ and $dQ/dt=0$ in equation~(\ref{eqn:m99}):
\begin{equation}
    \dot{n}_{\rm ion, crit} \geq C \langle n_{\rm H}\rangle^2 \alpha_{\rm HII}.
\end{equation}
We note that this expression can be re-stated as a criterion that reionization \emph{progresses} at a given value of the ionized fraction (e.g.~\citealt{Cullen24}),
\begin{equation}
    \dot{n}_{\rm ion,crit}(Q) \geq Q C \langle n_{\rm H} \rangle^2 \alpha_{\rm HII} = Q\times \dot{n}_{\rm ion,crit}.
\end{equation}
i.e.~for the ionized fraction to increase with time, the number of new ionizations must be larger than the number of recombinations within the ionized phase of the IGM.

In Figure~\ref{fig:nioncrit}, we compare the critical values of ionizing photon emissivity for ionized fractions of 25\%--100\% at $z=6$--$8$ with the corresponding emissivity calculated from the UV LFs versus the UV magnitude integration limit. As in Figure~\ref{fig:history1}, we assume $f_{\rm esc}\xi_{\rm ion} = 10^{24.8}$\,erg/Hz. Under this assumption, galaxies at $z=6$ can maintain reionization provided that ionizing photons escape from galaxies as faint as $M_{\rm UV}\sim-14$, while at $z=7$ and $z=8$ this would only be sufficient to continue reionizing the universe at ionized fractions of $50\%$ and $25\%$, respectively.

\newpage

\section{Summary \& Conclusion}

In this work, we have explored the implications of the short mean free path of ionizing photons at $z\approx6$ \citep{Becker21,Zhu23} for the recombination rate in the intergalactic medium as a whole, quantified by the clumping factor $C$. We first build an analytic connection between the mean free path and the recombination rate with minimal assumptions. The number density of ionizing photons in the optically-thin IGM can be derived from the hydrogen photoionization rate $\Gamma_{\rm HI}$ measured from the Ly$\alpha$ forest, and the rate at which these photons perform an ionization can be derived from the photon lifetime implied by the typical distance they travel before being absorbed, i.e. the mean free path. A global value for $C$ can then be estimated by comparing this rate to the recombination rate expected for a uniform IGM at the cosmic mean density.

We find a characteristic value of $C\approx12$ at $z=5$--$6$ that is well in excess of the $C=3$ assumption commonly made in the literature based on cosmological radiation-hydrodynamics simulations. Surprisingly, this elevated value persists to later times, consistent with non-evolution down to $z\sim2.5$. We tentatively attribute our higher value of $C$ to the way in which simulation analyses explicitly neglect dense gas within galaxy halos. While such an exclusion appears necessary to avoid double-counting the gas responsible for the galactic escape fraction, it ignores the fact that this dense gas can still absorb \emph{external} photons streaming through the IGM, and thus play an important role in determining the total budget of recombinations. 

Compared to the typical assumption of $C=3$, we find that late-ending reionization histories with $C=12$ require roughly twice as many ionizing photons to complete the process at $z\lesssim6$. However, recent observations of the ionizing efficiency of $z>6$ galaxies from JWST (e.g.~\citealt{Simmonds24}) and scaling relations for the ionizing escape fraction from low-redshift Lyman continuum leakers (e.g.~\citealt{Chisholm22}) imply a tremendous \emph{excess} in the ionizing photon budget \citep{Munoz24}. Due to the difference in our assumed recombination coefficient, the recombination rate in our fiducial model with $C=12$ is comparable to that of the $C=20$ model explored by \citet{Munoz24} in which reionization still ends quite early at $z\sim7.5$. 

We note also that the clumping factor may not provide a complete picture of the number of photons required to complete the reionization process. As shown in \citetalias{Davies21}, the fact that the ionizing sources and neutral islands are physically offset from one another implies a large degree of attenuation, requiring $\sim6$ photons per baryon to reach a neutral fraction of $x_{\rm HI}\sim10\%$ at $z\sim6$; about a factor of two higher than our fiducial model here with $C=12$. It is possible that both a large recombination rate and a consideration of the physical offset are required to reconcile the copious ionizing photon production of the first galaxies with current constraints on the reionization history.

\begin{acknowledgements}
The manuscript was completed following productive discussions with Girish Kulkarni, Laura Keating, Anson D'Aloisio, and Christopher Cain at the NORDITA workshop programme ``Cosmic Dawn at High Latitudes''. 

SEIB is supported by the Deutsche Forschungsgemeinschaft (DFG) under Emmy Noether grant number BO 5771/1-1.
\end{acknowledgements}

\appendix
\vspace{-1.5em}
\section{Systematic variation with frequency dependence assumptions}\label{sec:freq}

The derivation of the effective global clumping factor in \S~\ref{sec:clump} relies on two assumptions of frequency dependence: the spectral index of the UV background intensity ($J_\nu\propto\nu^{-\alpha_b}$) and the mean free path ($\lambda_\nu\propto\nu^{\alpha_\lambda}$). These two indices are not completely independent -- at higher redshifts, the connection can be described via the absorption-limited approximation $J_\nu \propto \epsilon_\nu \lambda_\nu$ \citep{MW03}, where $\epsilon_\nu\propto\nu^{-\alpha_\epsilon}$ is the specific ionizing emissivity of the sources. There is a further frequency dependence in our conversion from the constraints on $\Gamma_{\rm HI}$ from the Ly$\alpha$ forest to $J_{\rm HI}$, which involves an integral over the hydrogen photoionization cross-section (equation~\ref{eqn:GJ}). Approximating the cross-section as $\sigma_{\rm HI}\propto\nu^{-3}$, the combined frequency dependent term can be written as
\begin{equation}
    C(\alpha_b,\alpha_\lambda) \propto \frac{\alpha_b+3}{1-4^{\alpha_b+3}} \frac{1-4^{\alpha_b+\alpha_\lambda}}{\alpha_b+\alpha_\lambda},
\end{equation}
although in practice we integrate over the full form of the photoionization cross-section from \citet{Verner96}.

\begin{figure}[h]
\begin{center}
\resizebox{8.5cm}{!}{\includegraphics[trim={1em 1em 1em 1em},clip]{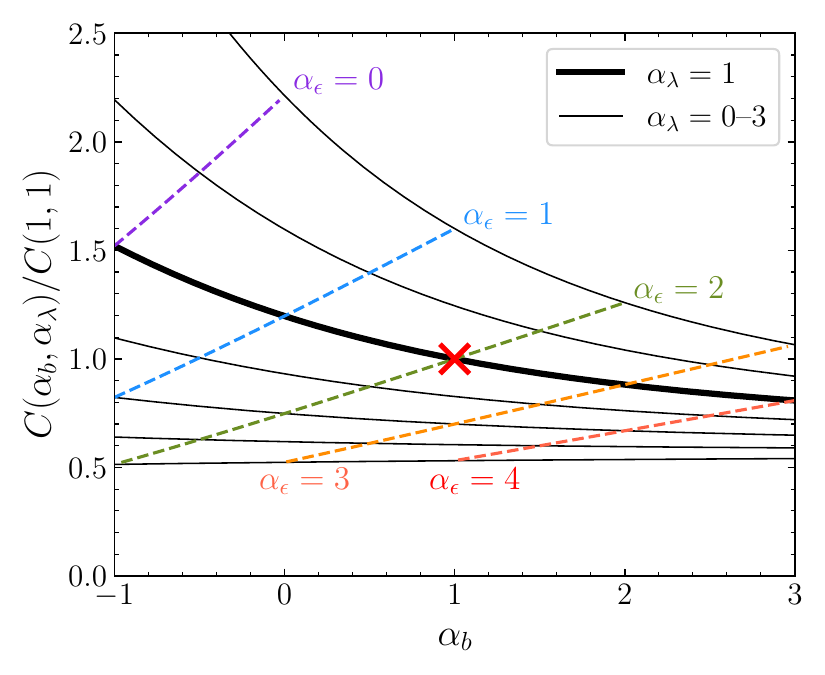}}\\
\end{center}
\caption{Effect of different frequency dependence assumptions on the estimated clumping factor relative to our fiducial choice of $\alpha_b=1$ and $\alpha_\lambda=1$ (red cross). The thick curve shows the variation with $\alpha_b$ at fixed $\alpha_\lambda=1$, with thin curves showing $\alpha_\lambda=0$--$3$ (top to bottom) in steps of $0.5$. The dashed colored lines show fixed values of $\alpha_\epsilon=\alpha_b+\alpha_\lambda$, where our fiducial choice corresponds to $\alpha_\epsilon=2$.}
\label{fig:freq}
\end{figure}

In Figure~\ref{fig:freq}, we show the full frequency dependence of the clumping factor calculation as a function of $\alpha_b$ for different values of $\alpha_\lambda$, taking into account the frequency dependencies in equation~(\ref{eqn:Cfid}) and equation~(\ref{eqn:GJ}). Constant values of $\alpha_\epsilon = \alpha_b+\alpha_\lambda$ are indicated, showing the interplay between these three quantities and the estimated clumping factor. Values of $C$ as low as half of our fiducial estimates are possible only if the mean free path increases sharply with frequency. 

We note that the column density distribution assumed in modern cosmological radiative transfer calculations is not a single power-law, but instead described by a piecewise series of power-laws surrounding the most relevant \ion{H}{1} column densities within $\pm2$ dex of the $N_{\rm LLS}=10^{17.2}$\,cm$^{-2}$ \citep{HM12,Puchwein19,FG20}. In these models, the slope of the distribution at $N_{\rm LLS}$ can be very steep; for example, \citet{Puchwein19} adopt $f(N)\propto N^{-1.95}$ in the range $10^{16}$--$10^{18}$\,cm$^{-2}$. Analytically, this should result in an extremely steep dependence of the mean free path with frequency, with $\alpha_\lambda\sim3$. Instead, we find that the (no longer power-law) frequency dependence arising from integrating over their column density distribution model at $z=6$ is better approximated by $\alpha_\lambda\sim1.6$ close to $\nu_{\rm HI}$, turning over to $\alpha_\lambda<1$ above $2\nu_{\rm HI}$. Self-consistently adopting this parameterization for $\lambda_\nu$ in our estimation of the clumping factor in equation~(\ref{eqn:Cfid}), including the resulting non-power-law shape of $J_\nu$, would reduce our $C$ from 12 to 10. However, a modest hardening of the source spectrum to $\alpha_\epsilon=1.2$ would return $C$ to 12.

\bibliographystyle{aasjournal}
 \newcommand{\noop}[1]{}

\end{document}